\documentclass[aps,prc,showpacs,twocolumn,preprintnumbers,superscriptaddress,nofootinbib,floatfix,10pt]{revtex4-1}
\usepackage{amsfonts,amssymb,stmaryrd,latexsym,amsmath,braket}
\usepackage{graphicx,subfigure}
\usepackage{comment}
\usepackage{times}
\usepackage{slashed}
\usepackage{bm}
\usepackage{braket}
\usepackage{chapterbib}
\usepackage{color}
\usepackage{xcolor}
\usepackage[export]{adjustbox}

\usepackage{hyperref}  

\usepackage{tabularx} 

\usepackage{multirow}
\usepackage{makecell}
\usepackage{booktabs}
\usepackage{array}
\newcommand{\PreserveBackslash}[1]{\let\temp=\\#1\let\\=\temp}
\newcolumntype{C}[1]{>{\PreserveBackslash\centering}p{#1}}
\newcolumntype{R}[1]{>{\PreserveBackslash\raggedleft}p{#1}}
\newcolumntype{L}[1]{>{\PreserveBackslash\raggedright}p{#1}}

\makeatletter
\let\saved@includegraphics\includegraphics
\AtBeginDocument{\let\includegraphics\saved@includegraphics}
\makeatother

\begin{document}

\title{Proton radioactivity in deformed nuclei with microscopic optical potential: A novel angular-dependent emission mechanism in the nanosecond-lived $^{149}$Lu}


\author{Yin Fan}
\affiliation{School of Nuclear Science and Technology, University of South China, 421001 Hengyang, China}

\author{Sibo Wang}
\email{Corresponding author: sbwang@cqu.edu.cn}
\affiliation{Department of Physics and Chongqing Key Laboratory for Strongly Coupled Physics, Chongqing University, Chongqing 401331, China}
\affiliation{Department of Physics, Graduate School of Science, The University of Tokyo, Tokyo 113-0033, Japan}

\author{Xiao-Hua Li}
\email{Corresponding author: lxh@usc.edu.cn}
\affiliation{School of Nuclear Science and Technology, University of South China, 421001 Hengyang, China}
\affiliation{Key Laboratory of Advanced Nuclear Energy Design and Safety, Ministry of Education, 421001 Hengyang, China}
\affiliation{Cooperative Innovation Center for Nuclear Fuel Cycle Technology $\&$ Equipment, University of South China, 421001 Hengyang, China}
\affiliation{National Exemplary Base for International Sci $\&$ Tech. Collaboration of Nuclear Energy and Nuclear Safety, University of South China, 421001 Hengyang, China}

\author{Haozhao Liang}
\email{Corresponding author: haozhao.liang@phys.s.u-tokyo.ac.jp}
\affiliation{Department of Physics, Graduate School of Science, The University of Tokyo, Tokyo 113-0033, Japan}
\affiliation{Quark Nuclear Science Institute, The University of Tokyo, Tokyo 113-0033, Japan}
\affiliation{RIKEN Center for Interdisciplinary Theoretical and Mathematical Sciences, Wako 351-0198, Japan}

\begin{abstract}

We present a theoretical description of proton radioactivity in $^{149}$Lu, the most oblate deformed proton emitter known, by combining a deformed microscopic optical potential derived from \textit{ab initio} nuclear matter calculations with the Wentzel–Kramers–Brillouin penetration probabilities and the assault frequency of the emitted proton estimated through a new harmonic‑oscillator‑inspired scheme. We predict a novel angular-dependent phenomenon unprecedented in spherical proton emitters: the disappearance of classically allowed regions at small polar angles $(\theta\leq 21^\circ)$. Our framework yields a half-life $T_{1/2}=467^{+143}_{-108}$\ ns for $^{149}$Lu, in excellent agreement within uncertainties with the experimental value $450^{+170}_{-100}$ ns. Deformation analysis rigorously excludes configurations with $|\beta_2|\geq 0.32$. Extensions to $^{150, 151}$Lu and their isomers also achieve excellent agreement with experimental half-life data. We further predict $^{148}$Lu as another highly oblate $(\beta_2 = -0.166)$ proton emitter with a half-life $T_{1/2}=4.42$\ ns. This work validates deformed microscopic optical potentials as a robust predictive tool for drip-line proton emitters and provides quantitative evidence for deformation effects in exotic decays.

\end{abstract}

\maketitle

\date{today}  


\textit{Introduction}. Proton radioactivity, the spontaneous emission of protons from proton-rich nuclei far from the $\beta$-stability, provides a unique probe of nuclear structure and dynamics beyond the drip line. First observed in 1970 through the decay of an excited isomer in $^{53}$Co \cite{1970-Jackson-PLB, 1970-Cerny-PLB}, this rare process was later confirmed in nuclear ground states ($^{151}$Lu and $^{147}$Tm) in 1982 \cite{1982-Hofmann-ZPA, 1982-Klepper-ZPA}. Recent advances in radioactive beam facilities have expanded the observation of proton emission events to over sixty cases \cite{2017-WangF-PLB, 2019-Auranen-PLB, 2021-Doherty-PRL, 2022-Auranen-PRL, 2022-ZhangW-ComPhys., 2025-Kokkonen-NatCom, 2025-XuXD-PRL}, yet predictive models for their decay properties remain challenged by nuclear deformation and continuum coupling \cite{2006-Delion-PR, 2008-Blank-PPNP, 2012-Pfutzner-RMP, 2019-QiC-PPNP}.

Recently, $^{149}$Lu was identified as a ground-state proton emitter with a record-high decay energy $Q_p = 1920(20)$\ keV and a nanosecond-scale half-life $T_{1/2}=450^{+170}_{-100}$ ns \cite{2022-Auranen-PRL}. Nonadiabatic quasiparticle calculations within phenomenological particle-rotor model suggest this nucleus may exhibit the strongest oblate deformation among known proton emitters \cite{2003-Fiorin-PRC}. In parallel, nuclear energy density functional (EDF) studies incorporating quantum tunneling rule out the possibility of a prolate quadratic deformation \cite{2023-XiaoY-PLB, 2024-LuQ-PLB}. However, their dependence on phenomenological interactions calibrated solely to $\beta$-stable nuclei invalidates the results for such exotic system, leaving theoretical predictions from realistic nucleon-nucleon (NN) interactions for this shortest directly measured half-life elusive.

In the aforementioned studies of $^{149}$Lu using nuclear EDF \cite{2023-XiaoY-PLB, 2024-LuQ-PLB}, the linear term of scalar and vector potentials $U_\text{S} + U_0$ was used as the nuclear potential to describe the proton emission phenomenon. This widely used approximation omits the quadratic contribution inherent in the Schr\"odinger equivalent potential $U_\text{SEP}$. As will be explained later, this omission artificially deepens the potential barrier, which not only affects the predicted half-life, but also obscures key deformation-sensitive emission dynamics in the quantum tunneling process.

In this letter, we integrate a deformed microscopic optical potential, developed from realistic NN interactions within the relativistic Brueckner-Hartree-Fock (RBHF) theory in the full Dirac space \cite{2021-WangSB-PRC, 2024-QinPP-PRC}, into the Wentzel-Kramers-Brillouin (WKB) framework for proton decay to calculate the half-life of $^{149}$Lu. Our method predicts a novel angular-dependent emission mechanism, which is further validated by successfully reproducing the half-lives of $^{150,151}$Lu and their isomers, and subsequently applied to predict the half-life of $^{148}$Lu. Throughout this work, we employ natural units where $\hbar=c=1$.


\textit{Half-life of proton radioactivity}. Proton emission is predominantly a quantum tunneling process \cite{2006-Delion-PR}. Its half-life is given by $T_{1/2} = \ln 2/(S_p \nu_0 P)$, where $S_p$ is the spectroscopic factor, $\nu_0$ is the assault frequency, and $P$ is the crucial penetration probability quantifying the likelihood of tunneling through the potential barrier.

The penetration probability $P$ is calculated within the WKB approximation
\begin{equation}\label{eq-P1}
  P = \exp \left[ -2 \int^{r_3}_{r_2} \sqrt{2\mu |Q_p - V(r)|} dr\right].
\end{equation}
Here, $V(r)$ is the total potential barrier hindering the emission of the proton, $r$ is the center-of-mass distance between the emitted proton and the daughter nucleus, $r_2$ and $r_3$ are the second and third classical turning points of $V(r)$, $\mu$ is the reduced mass of the proton-daughter system, and $Q_p$ is the decay energy. The potential barrier comprises three components: the nuclear potential $V_\text{N}(r)$, the Coulomb potential $V_\text{C}(r)$ and the centrifugal potential $V_l(r)$
\begin{equation}\label{eq-VNCl}
  V(r) = V_\text{N}(r) + V_\text{C}(r) + V_l(r).
\end{equation}
Notice that $P$ is highly sensitive to both $Q_p$ and the angular momentum $l$ embedded in $V_l(r)$, making its accurate determination essential.

\begin{figure}[htbp]
  \centering
  \includegraphics[width=8.5cm]{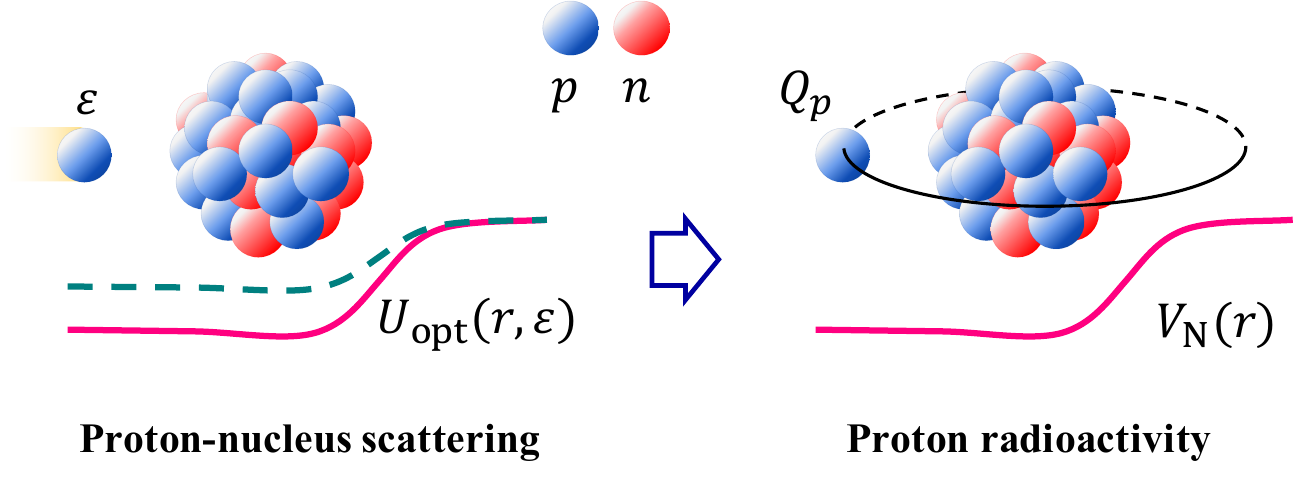}
  \caption{Sketch of assigning nuclear potential $V_\text{N}(r)$ in proton emission as the real part (pink solid) of proton-nucleus optical potential $U_\text{opt}(r,\varepsilon)$ with the imaginary part (cyan dashed) disregarded.}
  \label{Fig1}
\end{figure}

$V_\text{N}(r)$ describes the mean field experienced by the emitting proton due to its nuclear interaction with all the nucleons in the daughter. Obtaining accurate solutions for finite nuclear many-body systems based on realistic NN interactions is undoubtedly challenging. Previous studies predominantly employ phenomenological interactions containing adjustable parameters. Here, we assign $V_\text{N}$ as the real part of a microscopic nucleon-nucleus optical potential $U_\text{opt}$ which can be constructed from \textit{ab initio} solutions for homogeneous nuclear matter
\begin{equation}
  V_\text{N}(r) = \text{Re}\ U_\text{opt}(r,\varepsilon)|_{\varepsilon = Q_p}.
\end{equation}
Here $\varepsilon$ is the single-particle energy. This assignment seeks to establish a consistent theoretical treatment of both scattering and quasi-bound states. As shown in Fig.~\ref{Fig1}, the imaginary part of the optical potential is disregarded since no inelastic channels exist. Intuitively, this methodology is exceptionally well-suited for proton emission, owing to the characteristically weak binding of the proton-daughter system.

\textit{Microscopic optical potential RBOM}. The optical model provides a key tool for studying nucleon-nucleus scattering by reducing the complicated interactions between an incident nucleon and a target nucleus to a complex mean field, the optical potential. With the significant expansion of the nuclear landscape, there is a growing demand for microscopic optical potentials derived from realistic NN interactions \cite{2001-Deb-PRL, 2008-Quaglioni-PRL, 2012-Hagen-PRC, 2016-Vorabbi-PRC, 2016-Lynn-PRL, 2016-XuRR-PRC, 2017-Rotureau-PRC, 2019-Idini-PRL, 2019-Furumoto-PRC, 2021-Whitehead-PRL, 2023-KuangY-EPJA, 2024-QinPP-PRC}.

Instead of solving the full $A+1$-body problem, the improved local density approximation (LDA) provides an effective method to develop a microscopic optical model by equating the optical potential $U_\text{opt}$ to the nuclear matter single-particle potential $U_{\text{NM}}$ \cite{1959-Bell-PRL} augmented by a Gaussian convolution \cite{2016-XuRR-PRC, 2021-Whitehead-PRL, 2024-QinPP-PRC}
\begin{equation}\label{eq-ULDA}
    U_\text{opt}(r,\varepsilon) 
    = \int \frac{ \mathrm{d}^3r'}{(t\sqrt{\pi})^{3}}\ U_{\text{NM}}(\varepsilon, \rho(r'), \alpha(r'))\ e^{-\frac{|\bm{r}-\bm{r}'|^2}{t^2}}.
\end{equation}
Here, the density $\rho = \rho_n + \rho_p$ (where $\rho_n$ and $\rho_p$ are the neutron and proton densities, respectively) and isospin asymmetry $\alpha = (\rho_n-\rho_p)/\rho$ are determined by the target nucleus density profile, and $t$ parametrizes the effective range of NN interaction.

Inspired by the success of Dirac phenomenology in nucleon-nucleus scattering \cite{1979-Arnold-PLB, 1979-Arnold-PRC} and that of RBHF theory in nuclear matter saturation \cite{1990-Brockmann-PRC, 2021-WangSB-PRC}, we developed a RBHF-based microscopic optical model (RBOM) using the improved LDA method \cite{2024-QinPP-PRC}. The single-particle potentials in nuclear matter are self-consistently obtained in the full Dirac space \cite{2021-WangSB-PRC, 2022-WangSB-PRC}, whereas the target densities are calculated with the covariant EDF theory \cite{2015-MengJ-book}. Within the relativistic framework, the central term of the RBOM potential is obtained through the Schr\"odinger equivalent potential $U_\text{SEP}$~\cite{1980-Jaminon-PhysRevC.22.2027} 
\begin{equation}\label{eqUopt}
   U_\text{NM} = U_\text{SEP} \equiv U_\text{S} + \frac{E}{M}U_0 + \frac{1}{2M} (U^2_\text{S} - U^2_0),
\end{equation}
where $U_\text{S}$ and $U_0$ are the scalar and vector potentials, respectively, $M$ is the rest mass of a nucleon, and $E$ is the total single-particle energy containing rest mass. The RBOM potential successfully reproduces proton scattering observables for $^{208}$Pb, $^{120}$Sn, $^{90}$Zr, $^{48}$Ca, and $^{40}$Ca up to 200 MeV \cite{2024-QinPP-PRC}, exhibiting remarkable agreement with experimental data.

For the axially deformed $^{149}$Lu, we generalize the spherical RBOM potential using the improved LDA in two-dimensional space, where the Schr\"odinger equivalent potential $U_\text{SEP}(\varepsilon, \theta, \rho(r,\theta), \alpha(r,\theta))$ becomes angular dependent, with $\theta$ being the orientation angle of the proton with respect to the symmetric axis of the daughter nucleus. In this work, we obtain two-dimensional nucleon densities using the deformed relativistic Hartree-Bogoliubov theory in continuum (DRHBc) \cite{2020-ZhangKY-PRC} with the PC-PK1 density functional \cite{2010-ZhaoPW-PRC}. The DRHBc theory has demonstrated its remarkable ability in providing satisfactory description of the ground-state properties with powerful explorations \cite{2021-ZhangKY-PRCL}, thanks to the self-consistent consideration of the nuclear superfluidity, deformation, and continuum effects. The deformation also brings angular-dependent penetration probability
\begin{equation}\label{eq-P2}
  P_\theta = \exp\left[-2\int_{r_2(\theta)}^{r_3(\theta)}\sqrt{2\mu|Q_p - V(r,\theta)|}\,{\rm d}r\right].
\end{equation}
Neglecting angular correlations in the proton emission process, the total penetration probability is averaged over orientations as $P = \frac{1}{2} \int_{0}^{\pi} P_{\theta} \sin \theta \text{d}\theta$.


\begin{figure*}[htbp]
  \centering
  \includegraphics[width=14cm]{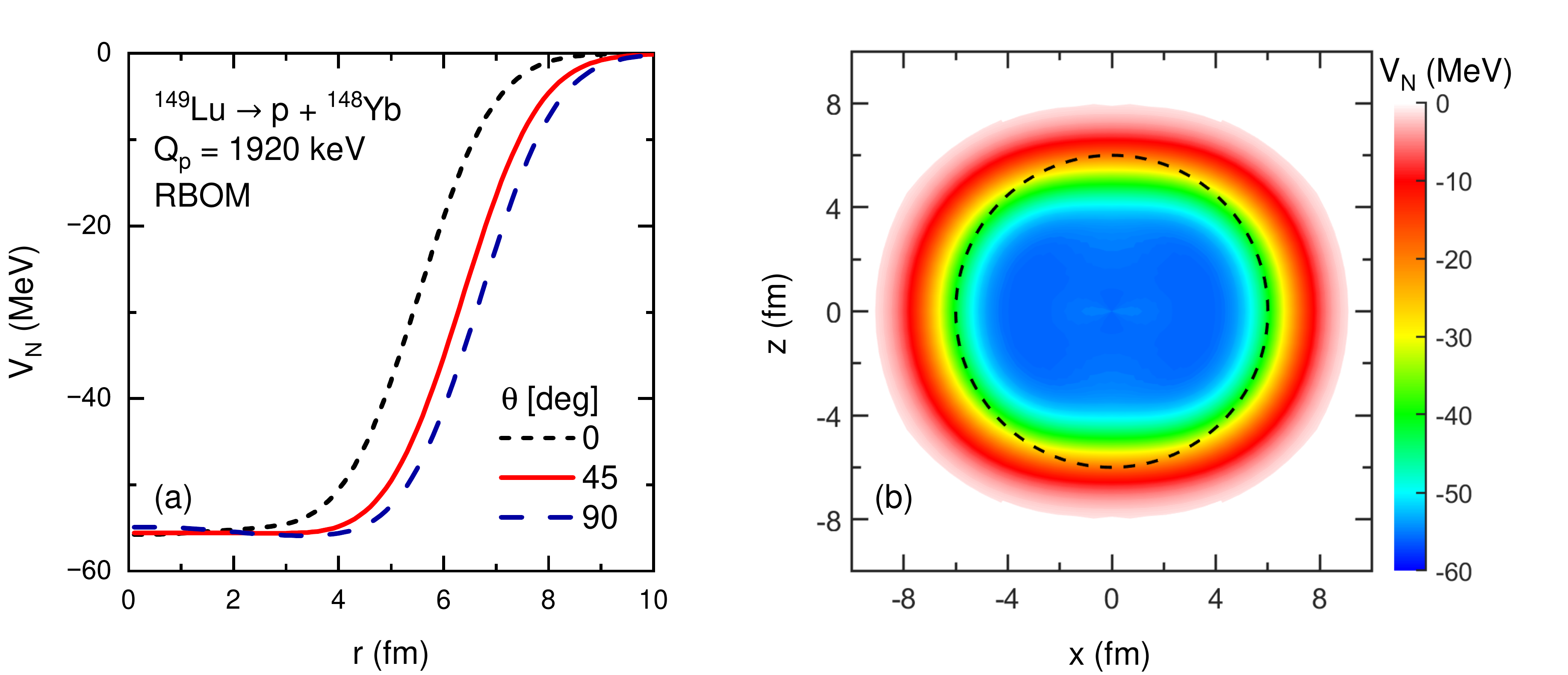}
  \caption{Nuclear potential $V_\text{N}(r,\theta)$ for proton emission of $^{149}$Lu. (a) Radial dependence at $\theta =0^\circ,\ 45^\circ$, and $90^\circ$ (b) Contour plot in the $xz$ plane. A dashed circle in panel (b) is drawn to guide the eye.}
  \label{Fig2}
\end{figure*}

\textit{Results and discussion}. Figure \ref{Fig2} depicts the nuclear potential $V_\text{N}(r,\theta)$ for proton emission of $^{149}$Lu obtained from the two-dimensional microscopic optical potentials. Panel (a) shows its radial dependence at $\theta = 0,\ 45^\circ$, and $90^\circ$, revealing a Woods-Saxon-like shape, ranging from approximately $-56$ MeV at $r=0$ to zero at $r\approx10$ fm. In comparison, the maximal value of the imaginary part of the optical potential is at most $-6$ MeV, supporting its neglect. The $\theta$-dependence is most pronounced in the nuclear surface region of 4-8 fm, and the nuclear potential becomes shallower at smaller angles. Panel (b) displays the values of $V_\text{N}(r,\theta)$ in the $xz$ plane. Due to the oblate shape with $z$ as the symmetry axis, the nuclear potentials in the $xz$ and $yz$ planes are identical, and the results in the $xy$ plane are rotational invariant. 

\begin{figure}[!htbp]
  \centering
  \includegraphics[width=8.5cm]{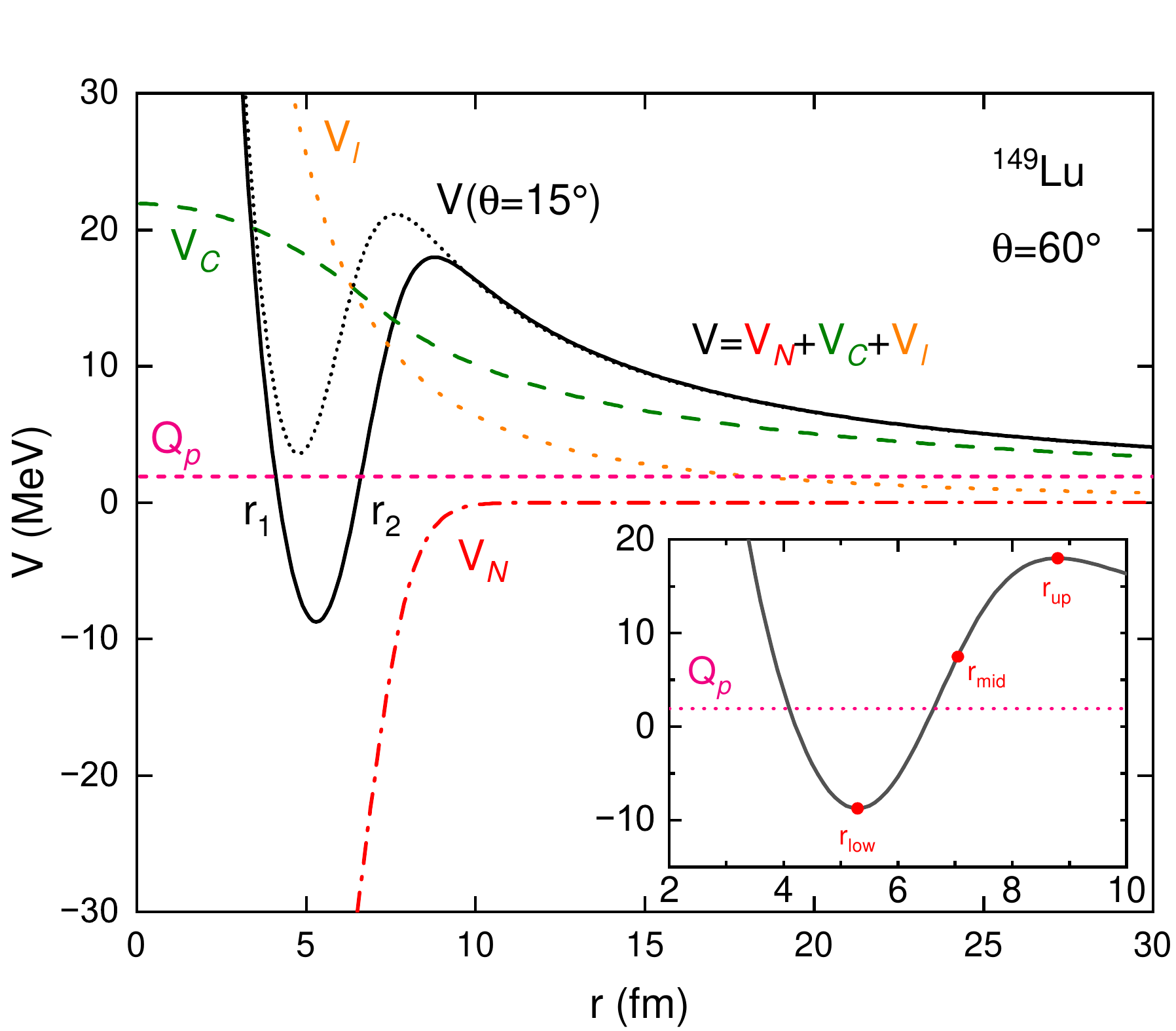}
  \caption{Proton-emission potential barrier for $^{149}$Lu at $\theta = 60^\circ$. Nuclear component ($V_\text{N}$, red dash-dotted) is derived from microscopic optical potential; Coulomb component ($V_\text{C}$, green dashed) is calculated via the double-folding method using daughter-nuclei deformation; Centrifugal component ($V_l$, orange dotted) adopts the Langer modified form. The black short-dotted line shows the potential barrier at $\theta = 15^\circ$. Inset: Enlarged view of the potential near its minimum, with reference points (red dots) used to estimate the frequency $\omega(\theta)$.}
  \label{Fig3}
\end{figure}

Taking $\theta=60^\circ$ as an example, Fig.~\ref{Fig3} shows the nuclear, Coulomb, and centrifugal contributions to the potential barrier in Eq.~\eqref{eq-VNCl}, with an inset illustrating the potential near its minimum used to extract the frequency $\omega(\theta)$. The third classical turning point $r_3$, located far from the center (not shown), is primarily determined by the Coulomb potential, which is calculated in this work using the double-folding method \cite{2000-Takigawa-PhysRevC.61.044607,2003-Ismail-Phys.Lett.B}, taking the deformation $\beta_2 = -0.168$ of the daughter nucleus $^{148}$Yb as input. The innermost turning point $r_1$ is mainly determined by the centrifugal potential, which we take in the Langer modified form, $V_l(r) = (l+1/2)^2/(2\mu r^2)$ \cite{1995-Morehead-JMP} with $l = 5$ for $^{149}$Lu. In comparison, the second turning point $r_2$ near the nuclear surface exhibits the competition among three components.

We find that the potential barrier $V(r,\theta)$ for the proton can be well approximated by a harmonic oscillator potential with frequency $\omega(\theta)$ in the vicinity of the minimum. This motivates our estimation of the assault frequency $\nu_0$ of emitted proton, a quantity poorly constrained in proton emission studies, as $\nu_0(\theta) = \omega(\theta)/(2\pi)$. As illustrated in the inset of Fig.~\ref{Fig3}, $\omega(\theta)$ is extracted within this new harmonic-oscillator-inspired scheme using
\begin{equation}
  V(r_\text{mid},\theta) - V(r_\text{low},\theta)= \frac{1}{2}\mu \omega(\theta)^2 (r^2_\text{mid} - r^2_\text{low}),
\end{equation}
where the reference point $r_\text{mid}=(r_\text{low}+r_\text{up})/2$ is adopted.

\begin{figure*}[!htbp]
  \centering
  \includegraphics[width=14cm]{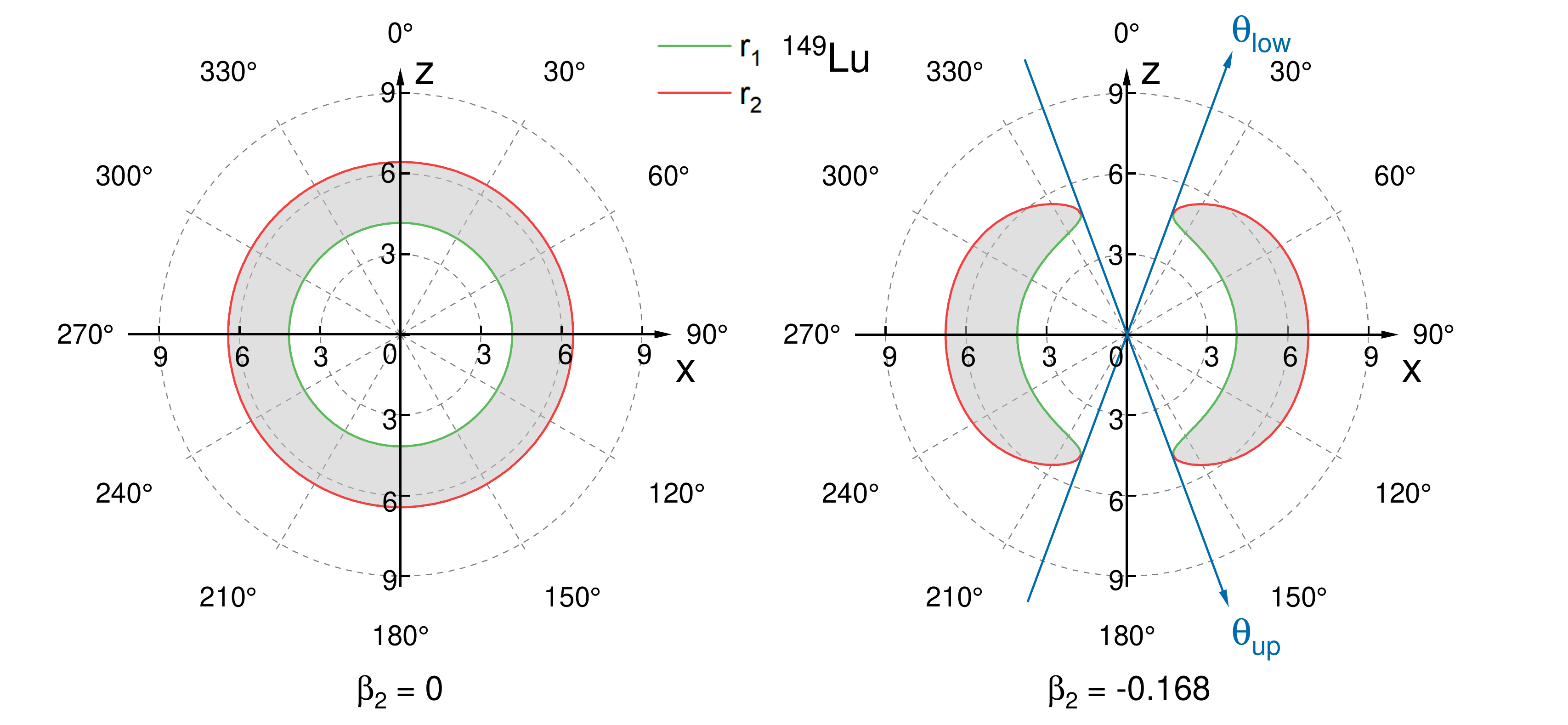}
  \caption{Classically allowed region (gray) enclosed by $r_1(\theta)$ (green) and $r_2(\theta)$ (red) in $xz$ plane for proton emission of $^{149}$Lu: spherical case (Left, $\beta_2=0$) vs. oblate deformation (Right, $\beta_2 = -0.168$). For the oblate case, protons are confined to $\theta\in[21^\circ,159^\circ]$ before tunneling, which drastically reduces the emission phase space compared to spherical systems.}
  \label{Fig4}
\end{figure*}

While the potential barrier for proton emission is rotationally invariant for spherical nuclei, deformed nuclei require explicit treatment of its angular dependence. As $\theta$ decreases from $90^\circ$, the barrier progressively shallows. Crucially, for $\theta\leq 21^\circ$, the barrier minimum exceeds $Q_p$, causing the turning points $r_1$ and $r_2$ to vanish, as demonstrated by the black short-dotted line at $\theta = 15^\circ$ in Fig.~\ref{Fig3}. Consequently, this complete suppression of the classically allowed region at these angles directly results in $P_\theta = 0$ . Figure~\ref{Fig4} quantifies the dramatic shrinkage of classically allowed regions (gray areas) bounded by $r_1(\theta)$ and $r_2(\theta)$, contrasting sharply with the rotationally invariant case of a spherical nucleus. Previous covariant EDF studies of $^{149}$Lu \cite{2023-XiaoY-PLB, 2024-LuQ-PLB} considered only the linear term $U_\text{S}+U_0$ in Eq.~\eqref{eqUopt}, omitting the quadratic contribution. This approximation deepens the potential barrier by approximately 20 MeV, allowing solutions for $r_1, r_2$ at all angles and thereby obscuring the predicted angular-dependent emission phenomenon. The superiority of $U_\text{SEP}$ over $U_\text{S}+U_0$ in relativistic proton emission calculations is demonstrated through Dirac-equation solutions for square-barrier tunneling and benchmark against Schr\"odinger-equation solutions. Therefore, in this work, we not only incorporate a microscopic optical potential into the description of proton radioactivity, but also go beyond previous approaches by addressing an aspect that has not been sufficiently explored, thereby uncovering a new physical mechanism in proton emission. Experimentally elucidating this novel angular-dependent emission mechanism, for instance through the angular anisotropy of protons emitted from polarized mother nuclei, would provide critical insights into its underlying physics.

\begin{table}[!htbp]
  \centering
  \caption{Half-life of proton emitter $^{149}$Lu predicted by microscopic optical potentials with the WKB approximation. Note that the central values for RBHF are obtained with $Q_p = 1920$\ keV and the parameter $t=1.3$\ fm in Eq.~\eqref{eq-ULDA}, while the uncertainties correspond to experimental measurements of $Q_p$.}
  \renewcommand{\arraystretch}{1.3}
  \setlength{\tabcolsep}{3.0mm}{
  \begin{tabular}{lllll}
  \hline
  \hline
  \multirow{2}{*}{$^{149}$Lu} & \multirow{2}{*}{Exp.~\cite{2022-Auranen-PRL}}  & \multicolumn{3}{l}{RBOM + WKB}  \\
  \cline{3-5}
            &     & Bonn A & Bonn B & Bonn C \\
  \hline
  $T_{1/2}$\ (ns) & 450$^{+170}_{-100}$ & 467$^{+143}_{-108}$ & 504$^{+154}_{-116}$ & 536$^{+174}_{-124}$ \\
  \hline 
  \hline 
  \end{tabular}}
  \label{Table1}
\end{table}

Table \ref{Table1} summarizes our theoretical proton-emission half-life for $^{149}$Lu, where $S_p = 0.813$ is obtained with DRHBc calculation as the nonoccupation probability of the corresponding orbital in the daughter nucleus. Our central prediction of $467\ \text{ns}$ using $Q_p = 1920$\ keV, $t = 1.3$\ fm and Bonn A, shows excellent agreement with the experiment value of $450$\ ns. We consider the following sources of theoretical uncertainty: (i) $Q_p$ variation ($1920(20)$\ keV), (ii) improved LDA range parameter ($t=1.25$ - $1.35$\ fm), and (iii) realistic NN interactions variants (Bonn A/B/C). $T_{1/2}$ is most sensitive to $Q_p$, while exhibiting only weak variation with $t$ ($\leq$ $5$\ ns). In the literature \cite{2016-XuRR-PRC, 2021-Whitehead-PRL}, the value of t was typically determined by fitting to experimental data, with optimal values lying in the range of $1.0$ - $1.4$\ fm. In this work, we adopt a typical value $t = 1.3$\ fm, as originally employed in the microscopic optical potentials for proton-nucleus scattering \cite{2024-QinPP-PRC}. Varying t between $1.25$ and $1.35$\ fm here is found to change the half-life of $^{149}$Lu by less than $5$\ ns. The difference among Bonn A, B, and C arises from the cutoff parameter in $\pi$N coupling, resulting in the weakest tensor force and thus the strongest central attraction for Bonn A~\cite{2024-WangSB-SciBull}. The increased depth of the nuclear potential reduces the integrated area between the barrier curve and $Q_p$ energy (Fig.~\ref{Fig3}), leading to higher penetration probability $P$ and correspondingly shorter half-life $T_{1/2}$.



Using the ground-state density of $^{148}$Yb with $\beta_2 = -0.168$, we constructed the deformed optical potential for proton emission of $^{149}$Lu. The near-identical potential energy curves (PECs) of the mother and daughter nuclei justify probing the deformation effects on half-life via the daughter's PEC. Assuming decay to the daughter's ground state (fixed $S_p$), the half-life variation stems solely from the potential barrier modifications through $P$ and $\nu_0$. Notably, the novel suppression of classically allowed regions occurs for $\beta_2\leq -0.1$ and $\beta_2\geq 0.3$, confirming its deformation-driven origin.

\begin{figure}[!htbp]
  \centering
  \includegraphics[width=8.5cm]{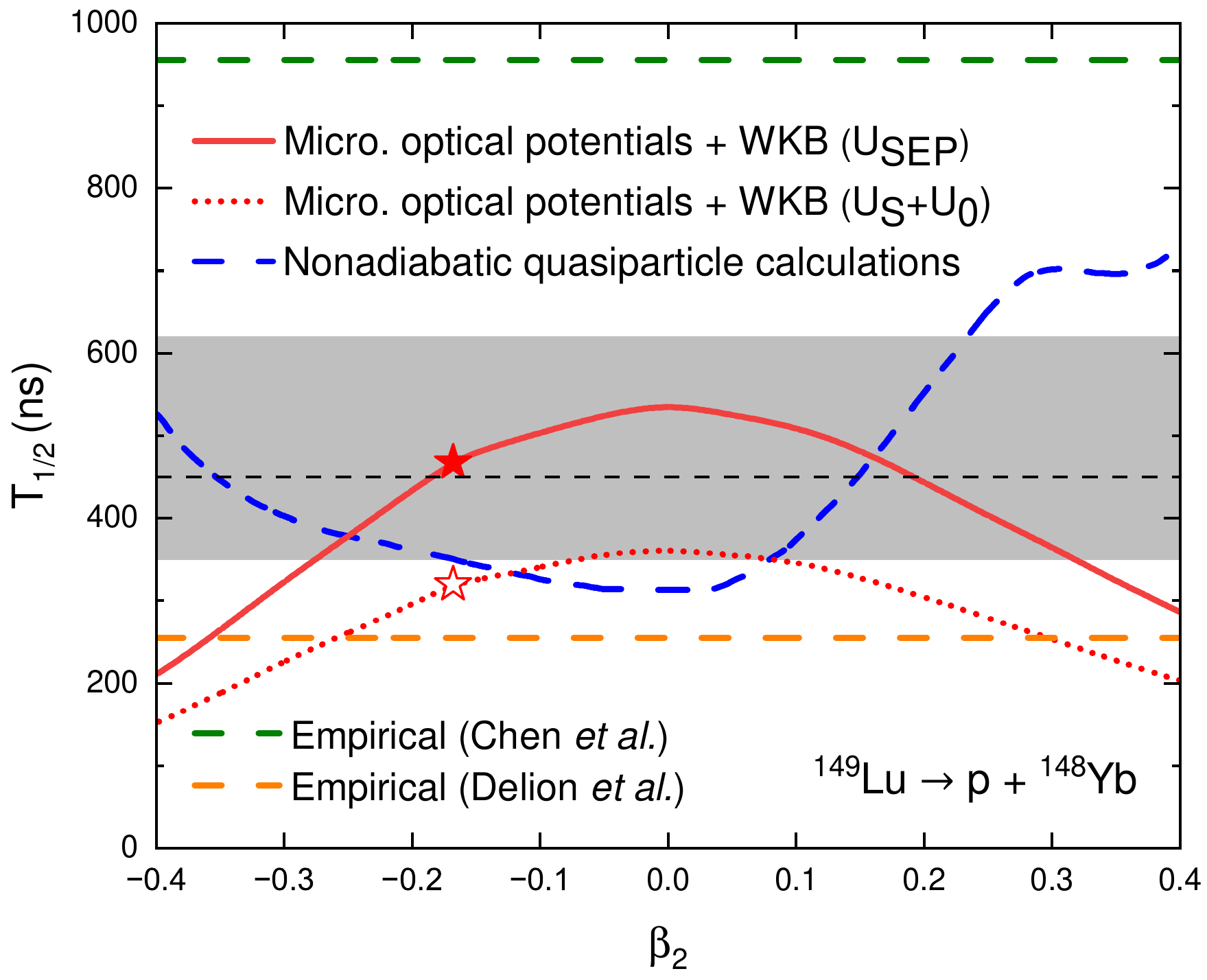}
  \caption{Half-life of $^{149}$Lu predicted by microscopic optical potentials with the WKB approximation (red solid) as a function of quadrupole deformation. Results with the Schr\"odinger equivalent potential $U_\text{SEP}$ replaced by $U_\text{S}+ U_0$ (red short-dotted), those from nonadiabatic quasiparticle calculations (blue dashed), and those with the empirical formulas from Chen \textit{et al.}~\cite{2019-ChenJL-EPJA} (green dashed) and Delion \textit{et al.}~\cite{2021-Delion-PhysRevC.103.054325} (orange dashed) are also shown for comparison. Experimental uncertainties are represented by the gray band, while their central value is indicated by the black dashed line. The stars mark the oblate ground state with $\beta_2 = -0.168$.}
  \label{Fig5}
\end{figure}

As shown in Fig.~\ref{Fig5}, our calculations agree with the experimental half-life (gray band) for $-0.28\leq\beta_2\leq 0.32$. The prediction for the ground-state deformation ($\beta_2 = -0.168$, red star) matches the central experimental value within 4\%. Larger oblate and prolate deformations with $|\beta_2|\geq 0.32$ are excluded by the data. Predictions from the empirical formulas of Chen \textit{et al.}~\cite{2019-ChenJL-EPJA} and Delion \textit{et al.}~\cite{2021-Delion-PhysRevC.103.054325} fall well outside the experimental uncertainty band, underscoring the significant challenge that $^{149}$Lu's anomalously short half-life poses to theoretical descriptions. Strikingly, replacing $U_\text{SEP}$ with $U_\text{S} + U_0$ shortens $T_{1/2}$ and worsens the agreement with experiment. Contrary to nonadiabatic quasiparticle calculations \cite{2022-Auranen-PRL}, our optical potential-based approach predicts maximum half-life at the spherical configuration. This fundamental discrepancy in deformation dependence highlights the need for future experimental analyses to incorporate findings from microscopic optical potential calculations. On the other hand, resolving this discrepancy requires further advances in comprehensive microscopic approaches to proton emission. 

\begin{table}[htbp]
  \centering
  \caption{Calculated proton radioactivity half-lives for $^{150,151}$Lu from ground and isomeric states using the Bonn A, $t=1.3$\ fm and experimental $Q_p$ values \cite{2021-Huang-CPC,2009-Singh-NDS}, in comparison to experimental data \cite{2008-Blank-PPNP,2009-Singh-NDS,2021-Kondev-CPC}. Orbital angular momentum quantum numbers are taken from Ref.~\cite{2021-Kondev-CPC}, and deformation parameters $\beta_2$ are derived from the DRHBc calculations with PC-PK1. Theoretical $Q_p$ value from \cite{2022-NiuZM-PhysRevC.106.L021303} (with underline) is used for $^{148}$Lu prediction.}
  \renewcommand{\arraystretch}{1.3}
  \setlength{\tabcolsep}{2.mm}{
  \begin{tabular}{rccccc}
  \hline 
  \hline 
    {Nucleus} & $l$ & $\beta_2$ & $Q_p$\ (keV) & $T_{1/2}^\text{Exp.}$ & $T_{1/2}^\text{Cal.}$\\
  \hline
  $^{150}$Lu  & 5 & -0.153 & 1270 & 45(3) ms          & 51.12 ms  \\
  $^{150m}$Lu & 2 &  0.142 & 1291 & 40(7) $\mu$s      & 18.50 $\mu$s \\
  $^{151}$Lu  & 5 & -0.145 & 1241 & 127.1(18) ms      & 104.33 ms  \\
  $^{151m}$Lu & 2 &  0.117 & 1289 & 16.0(0.5) $\mu$s  & 9.57 $\mu$s \\
  $^{148}$Lu  & 5 & -0.166 & \underline{2334}    & /  & 4.42\ ns \\
  \hline 
  \hline 
  \end{tabular}}
  \label{Table2}
\end{table}

To further validate our approach, we calculate the proton emission half-lives of $^{150, 151}$Lu from both their ground and isomeric states. Here, we use the microscopic optical potential obtained with Bonn A and $t=1.3$\ fm. The vanishing of the first two turning points at small angles is also found for ground-state proton emission of $^{150, 151}$Lu. As shown in Table \ref{Table2}, our results exhibit good agreement with experimental data across four orders of magnitude, particularly for the ground-state proton emissions (within a factor of 1.2).

Finally, we extend our framework to predict the half-life of $^{148}$Lu with $\beta_2=-0.166$. The $Q_p$ value is adopted from a Bayesian machine learning mass model achieving 84 keV accuracy \cite{2022-NiuZM-PhysRevC.106.L021303}, yielding $Q_p=2.334\pm0.411$\ MeV. For the central $Q_p$ value, we predict $T_{1/2}=4.42$\ ns, which is two orders of magnitude shorter than the experimental 450 ns for $^{149}$Lu. This extreme disparity primarily reflects the higher decay energy. If experimentally confirmed, $^{148}$Lu would become the shortest-lived proton emitter known, providing a critical benchmark for drip-line theories.



\textit{Summary}. We develop a theoretical description of proton radioactivity integrating deformed microscopic optical potentials derived from realistic NN interactions, with WKB penetration probabilities and a newly proposed harmonic-oscillator-inspired assault frequency scheme. Crucially, we predict a novel angular-dependent emission mechanism, unprecedented in spherical proton emitters, where the classically allowed region of the proton is completely suppressed at small polar angles $(\theta\leq 21^\circ)$.

For $^{149}$Lu, our calculations reproduce the measured half-life with remarkable accuracy. Systematic deformation studies exclude configurations with $|\beta_2|>0.32$. The framework is successfully extended to neighboring $^{150,151}$Lu and predicts $T_{1/2}=4.42$\ ns for $^{148}$Lu. This work validates deformed microscopic optical potentials as a robust predictive tool for drip-line proton emitters and provides quantitative evidence for deformation effects in exotic decays.

Future theoretical investigations could fruitfully extend this work in two key directions. First, incorporating triaxial deformations in $^{149}$Lu \cite{2024-LuQ-PLB} would probe how $\gamma$-softness and axial symmetry breaking alter the angular dependence of the potential barrier, thereby modulating the angular-dependent emission pattern identified here. Second, replacing the meson-exchange Bonn potentials with relativistic chiral NN interactions \cite{2018-RenXL-CPC, 2022-LuJX-PRL} could strengthen the connection between the microscopic optical potential and fundamental quantum chromodynamics, while systematically accounting for relativistic kinematic and dynamical effects.


~~

\textit{Acknowledgments}. This work was partially supported by the National Natural Science Foundation of China (NSFC) under Grant Nos. 12175100, 11975132, 12205030, and 12575130.

\textit{Data Availability}. The data that support the findings of this article are not publicly available. The data are available from the authors upon reasonable request.

\bibliography{149Lu-main}

\end{document}